\documentclass[12pt]{article}
\usepackage{amsmath,amssymb,epsfig}

\def\K{{\mathcal K}}

\def\be{\begin{equation}}
\def\ee{\end{equation}}
\def\lp{\ell_P}
\def\R{{\mathcal R}}
\def\R{{\mathcal R}}

\def\K{{\mathcal K}}
\def\be{\begin{equation}}
\def\ee{\end{equation}}
\def\lp{\ell_P}

\def\beq{\begin{eqnarray}}\def\eeq{\end{eqnarray}}

\begin{document}
\title{Entanglement entropy from the holographic stress tensor}

\author{Arpan Bhattacharyya  and Aninda Sinha\\
\it $^1$Centre for High Energy Physics, Indian Institute of Science,  Bangalore, India.\\}

\maketitle
\vskip 2cm
\begin{abstract}{\small
We consider entanglement entropy in the context of gauge/gravity duality for conformal field theories in even dimensions. The holographic prescription due to Ryu and Takayanagi (RT) leads to an equation describing how the entangling surface extends into the bulk geometry. We show that setting to zero the time-time component of the Brown-York stress tensor evaluated on the co-dimension one entangling surface, leads to the same equation. By considering a spherical entangling surface as an example, we observe that Euclidean action methods in AdS/CFT  will lead to the RT area functional arising as a counterterm needed to regularize the stress tensor. We present arguments leading to a justification for the minimal area prescription.}
\end{abstract}

\newpage

Entanglement entropy of a local quantum field theory is a useful concept featuring in diverse areas ranging from black holes in general relativity  \cite{Bombelli:1986rw,Srednicki:1993im} to Fermi surfaces in condensed matter systems  \cite{Gioev:2006zz,Ogawa:2011bz}.
Entanglement entropy in conformal field theories in even $d$ dimensions \cite{ryu,Ryu:2006ef,odd} takes the form
\be\label{def}
S_{EE}=c_d \frac{l^{d-2}}{\epsilon^{d-2}}+O(\frac{l^{d-3}}{\epsilon^{d-3}})+a_d \log \frac{l}{\epsilon}+O((\frac{l}{\epsilon})^0)\,.
\ee
 Here $l$ is a length scale parametrizing the size of the entangling region and $\epsilon$ is a short-distance cutoff. The leading $l^{d-2}$ term gives the famous area law with a non-universal proportionality constant--when $d=2$ the leading term is the $\log$ term.
The coefficient of the $\log$ term is a universal quantity typically related to a function of the conformal anomalies in the theory \cite{ryu, adam, solod}. Entanglement entropy has also proved useful in quantifying the number of degrees of freedom in quantum field theories \cite{myersme, strip}. In the context of quantum field theories, a direct computation of entanglement entropy is hard and has been possible only
in very specific examples. Typically numerical techniques and the so-called replica trick are used \cite{calabrese}\,. Owing to its diverse applications \cite{myersnew}, it is of crucial importance to probe other computational tools available to us.

One useful computational prescription originally proposed by Ryu and Takayanagi (RT) comes from the gauge/gravity correspondence \cite{magoo}. The correspondence demands the existence of a duality between a quantum field theory in $d$ dimensions and a theory of 
gravity (possibly string theory) in one dimension higher. For computational purposes one typically uses Einstein gravity in a weakly curved anti-de Sitter (AdS) background which corresponds to a strongly coupled conformal field theory (CFT).
According to this prescription \cite{ryu}, in order to derive the holographic entanglement entropy for a $d$ dimensional quantum field theory,
one has to minimize the following entropy functional on a $d-1$ dimensional hypersurface (a co-dimension 2 surface),
\be \label{RT}
S=\frac{2\pi}{\ell^{d-1}_{P}} \int d^{d-1}x \sqrt{h}\,,
\ee
where $\ell_ P$ is the Planck length and $h$ is the induced metric on the hypersurface. The minimal surface extending into the bulk coincides with the entangling surface in the CFT at the AdS boundary. The gravity dual theory is simply Einstein gravity with a negative cosmological constant
\be \label{bulk}
I=-\frac{1}{\ell_{P}^{d-1}}\int d^{d+1}x \sqrt{g}\big [\frac{d(d-1)}{L^{2}}+ R \big]
\ee
where $g$ is the determinant of the bulk metric, $R$ is the scalar curvature for the bulk space time and $L$ is the AdS radius. The generalization of the RT prescription to a class of higher derivative theories of gravity called Lovelock theories has been proposed recently \cite{higher}. The hypersurface is a co-dimension 2 surface and extends into the extra dimension. The minimization of $S$ leads to an equation which gives the way the entangling surface extends into the bulk spacetime. In this paper, for definiteness we will consider $d=4$ and take the entangling surface to be either a sphere or a cylinder.

For the AdS$_{5}$  metric in Euclidean signature\,,
\be \label{met}
ds^{2}=\frac{L^{2}}{z^{2}}(dz^{2}+dt^{2}+\sum^{3}_{i=1} dx_{i}^{2})\,.\\
\ee
Here $z$ is the radial coordinate of the AdS space corresponding to the extra dimension. The field theory lives on the surface parametrized by $(t,x_i)$. The $z=0$ slice corresponds to the boundary of AdS  and according to the AdS/CFT dictionary corresponds to the ultraviolet (UV) regime of the field theory. 
Now we can choose, either $\sum^{3}_{i=1} dx_{i}^{2}=dr^{2}+r^{2}d\Omega_{2}^{2}$ corresponding to spherical coordinates for the boundary or $\sum^{3}_{i=1} dx_{i}^{2}=dv^{2}+dr^{2}+r^{2}d\phi^{2}$ corresponding to cylindrical coordinates for the boundary. Choosing the surface $t=0, r=f(z)$ one has to evaluate the entropy functional $S$, then find the equations of motion for $f(z)$ for the corresponding geometry of the entangling surface. Since we want the entangling surface in the field theory on the $z=0$ slice to be a sphere ($S^2$) or a cylinder ($R\times S^1$) we demand that $f(z)$ satisfies 
\be \label{fzero}
f(z)=f_{0}+f_1 z+ f_2 z^2+\cdots\,,
\ee
where $f_{0}$ gives the radius of the $S^2$ or the $S^1$. Let us review how this works. We first put $r=f(z), t=0$ in (\ref{met}). Then,\footnote{prime denotes derivative w.r.t $z$}
\be\label{ons}
S= \frac{2\pi}{\lp^{3}}\int d^{3}x\frac{L^{3} f(z)^n \sqrt{ \left(1+f'(z)^2\right)}}{z^3}\,.\\
\ee
where $n=1$ for the cylinder and $n=2$ for the sphere. The volume form  $d^3 x=\sin(\theta) dz d\theta d\phi$ for the sphere and $d^3x=dz dv d\phi$ for the cylinder. From here we get the following Euler-Lagrange equation for $f(z)$\,,
\be
\frac{L^3 \left[z f(z) f''(z)-\left(3 f(z) f'(z)+n z\right) \left(f'(z)^2+1\right)\right]}{\lp^{3}z^4  \left(f'(z)^2+1\right)^{3/2}}=0\,.\ee\\
Solving this equation as an expansion around $z=0$ fixes $f_{1},f_{2}$ in terms of $f_{0}$. It leads to
\be \label{sol1}
f_{1}=0,\quad f_{2}=-\frac{1}{4 f_{0}}\,,
\ee
for the cylinder.
For the sphere one can get an exact solution of the form
\be \label{sol2}
f(z)=\sqrt{f_{0}^{2}-z^{2}}\,.\\
\ee


When one evaluates the on-shell action eq.(\ref{ons}) and expands around $z=0$, then one gets a result exactly of the form in eq.(\ref{def}).
The RT prescription satisfies the strong subadditivity condition that entanglement entropy is known to satisfy and also passes some other nontrivial consistency checks \cite{Ryu:2006ef}.  However, attempted derivations \cite{fursaev} of this prescription are plagued with problems \cite{head}. For instance, the implementation of the Replica trick needs the introduction of a conical deficit in the spacetime that the field theory lives. This leads to an introduction of the conical singularities in the bulk and it is not known how to deal with such singularities consistently. Until recently, the only case where a derivation exists is in the case where the entangling surface is a sphere \cite{chm}--the situation has changed with a proposed derivation by Lewkowycz and Maldacena \cite{gengrav}. However, this derivation uses the replica trick in order to derive entanglement entropy. It is important to know if there are ways to derive entanglement entropy in holography that does not use the replica trick. As such it is important to explore the RT prescription to find clues that may shed light on an alternative way to a derivation. 




Let us begin by  calculating the stress tensor for the field theory living on a $r=f(z)$ co-dimension 1 surface using holography.
The Brown-York (holographic) stress tensor is given by \cite{stress}
\be \label{boundary}
T_{ab}=\frac{1}{\ell_{P}^{3}}(\K_{ab}-h_{ab}\,\K)+\frac{2}{\sqrt{h}}\frac{\delta S_{ct}}{\delta h^{ab}}.
\ee
where $\K_{ab}=e^{\alpha}_{a}e^{\delta}_{b} p^{\beta}_{\delta}\nabla_{\alpha} n_{\beta}$ is the extrinsic curvature, $\K$ is the trace of it,  $p^{\alpha}_{\gamma}=g^{\alpha}_{\gamma}-n^{\alpha}n_{\gamma}$ is the projection operator, $n_{\beta}$ is the normal for the surface $r=f(z)$, $\alpha , \beta , \gamma , \delta$  and $a,b$ are bulk and  boundary indices respectively. $e^{\alpha}_a$'s are the pullbacks.
 $S_{ct}$ is the counterterm action needed to regularize the stress tensor. We will address this contribution separately in a moment.  Conventionally, the stress tensor is evaluated on $z=\epsilon$ with $\epsilon\rightarrow 0$, corresponding to the UV of the field theory. However, we will compute the stress tensor on the slice $r=f(z)$. Furthermore, note that we will \underline{not} set $t=0$ when calculating the stress tensor, unlike the RT prescription. As a result we will be computing the tensor on a  4d slice. One could have equivalently chosen the slice $z=\rho(r)$. In this case for each fixed $z$, on the dual CFT side, we would have to consider the entangling surface at different RG scales. It may be possible to set up the RG equations leading to the entangling surface along these lines similar in spirit to \cite{tedren}. For this paper, we will compute on the $r=f(z)$ slice. Another point that we should emphasise is that we are assuming Dirichlet boundary conditions throughout. Neumann boundary conditions on the $r=f(z)$ slice appear in the considerations of Boundary Conformal Field Theory as in \cite{bcft}.

Now observe the following. Since the time direction is a direct product with the rest, the trace of the extrinsic curvature satisfies $^{(4)}K_a^a= ^{(4)}\!\!\!K_t^t+ ^{(3)}\!\!K_i^i$. Thus if we demand that $T_t^t=0=^{(4)}\!\!\!K_t^t-h_t^t \,{}^{(4)}\!K_a^a$ leads to $^{(3)}\!K_i^i=0$ which is the same as the minimal surface condition for the 3d slice used in the RT calculation. Of course, if we considered higher curvature gravity, the $T_t^t=0$ condition may be used to fix the surface $f(z)$. At this stage the stress tensor could still have divergences and we need to add a suitable $S_{ct}$ to remove them. If the $(\K_{tt}-h_{tt}\,\K)$ piece of the stress tensor is zero, then the counterterm should not affect this result. As such the relevant counterterm will turn out to live on the $t=0$ slice as we will explicitly show below.

 One can easily calculate $\K_{ab}$ and hence $T_{ab}$ using standard methods. We find
\begin{align}
\begin{split}
\K_{a}^{b}&=(K_{z}^{z},K_{t}^{t},K_{\theta}^{\theta},K_{\phi}^{\phi})\\&=\left(
 \frac{f'(z)^3+f'(z)-z f''(z)}{ L (f'(z)^2+1)^{3/2}},
 \frac{f'(z)}{L \sqrt{ \left(f'(z)^2+1\right)}} , \frac{((n-1) z + f(z)f'(z))}{L f(z) \sqrt{ \left(f'(z)^2+1\right)}}
, \frac{ \left(z+f(z) f'(z)\right)}{L f(z) \sqrt{ \left(f'(z)^2+1\right)}}\right)\,.
\end{split}
\end{align}
Here $n=2$ would correspond to a sphere and $n=1$ to a cylinder respectively. Thus we find that 
\be \label{tt}
T_{tt}= \frac{L  \left[z f(z) f''(z)-\left(3 f(z) f'(z)+n z\right) \left(f'(z)^2+1\right)\right]}{\lp^{3} z^2 f(z)  \left(f'(z)^2+1\right)^{3/2}}\,.
\ee
 Setting this to zero will lead to exactly the same equations  as the RT prescription as we argued. This also seems consistent with the fact that we are after all interested in the entanglement entropy of the ground state \cite{motiv}.

 We will now determine the counterterm needed to get a finite stress tensor. Let us focus on the sphere case.  Using the solution for $f(z)$ we can calculate the stress tensor as an expansion around the boundary $z=0$.
Let us start with the total Euclidean gravitational action. We will use the intuition gained from computing black hole entropy where the total action is evaluated by integrating from the horizon to infinity,
\be
I_{tot}=I_{bulk}+I_{GH}^{r=f(z)}+I_{ct}^{r=f(z)}+I_{GH}^{z=\epsilon}+I_{ct}^{z=\epsilon}\,.
\ee
 $I_{bulk}$ is given by eq.(\ref{bulk}) where the integration limits for $r$ go from $f(z)$ to $\Lambda$ with $\Lambda\rightarrow \infty$ is a radial cut-off, $I_{GH}^{r=f(z)}$ is the surface term for the $r=f(z)$ slice which is used in the calculation of eq.(\ref{tt}). $ I_{GH}^{z=\epsilon}, I_{ct}^{z=\epsilon}$ are the usual surface and counterterm \cite{stress} evaluated on the $z=\epsilon$ surface with $\epsilon\rightarrow 0$ corresponding to the usual AdS boundary.  We will rescale the time coordinate as $\hat t=\frac{t}{f_{0}}$. The stress tensor contribution coming from $r=f(z)$ slice is given by,
\begin{align}
\begin{split}
T_{ab}&=(T_{zz},T_{\hat t\hat t},T_{\theta\theta},T_{\phi\phi})\\
&=(\frac{L }{ f_{0} z},0,\frac{L f_{0}}{  z},\frac{L f_{0} \sin(\theta)^{2}}{  z})\,.\\
\end{split}
\end{align}
Note that there are $\frac{1}{z}$ divergences in the stress tensor in all components {\it except} $T_{tt}$ which is zero. Thus to regularize this stress tensor we will need 
 a  three dimensional counterterm on a co-dimension 2 surface. Remarkably, if we add the RT-functional in eq.(\ref{RT}) as $I_{ct}^{r=f(z)}= - S$, the stress-tensor becomes free of divergences.

Now let us evaluate various pieces of the total action as an expansion around $z=\epsilon$:
\be
I_{bulk}=-\frac{\pi^2 L^3}{2\lp^3}\left[\frac{16 f_{0} \left(f_{0}^3-\Lambda ^3\right)}{3  \epsilon ^4}-\frac{16  f_{0}^{2}}{ \epsilon^{2}}+8 \log(\frac{f_{0}}{\epsilon})+\cdots \right]\,.
\ee
\begin{align}
\begin{split}
I_{GH}^{r=f(z)}
&= -\frac{\pi^2 L^3}{\lp^3}\left[\frac{ 4 f_{0}^{2}}{\epsilon^{2} }-4 \log (\frac{f_{0}}{\epsilon})+\cdots\right]\,.
\end{split}
\end{align}
\be
I_{GH}^{z=\epsilon}+I_{ct}^{(z=\epsilon)}=\frac{ \pi^{2}    L^3}{\lp^{3}}\left[\frac{8 f_{0} \left(f_{0}^3-\Lambda ^3\right)}{3  \epsilon ^4}+\cdots\right]\,.\\
\ee
Lastly,
\be
I_{ct}^{r=f(z)}=-\frac{\pi^2 L^3}{\lp^3}\left[\frac{4 f_{0}^{2}}{\epsilon^{2} }-4\log(\frac{f_{0}}{\epsilon})+\cdots\right]\,.\\
\ee
Firstly notice that the $\log \epsilon$ pieces arising from $I_{bulk}$ and $I_{GH}^{r=f(z)}$ cancel.  How does one get entropy from here? When we have a black hole, we identify the periodicity of the Euclidean time as inverse temperature. In this case if we identify the periodicity of the time coordinate $t$ with the inverse Unruh temperature \cite{unruh, takatemp} $1/(2\pi f_0)$ which can be shown using independent arguments as in \cite{chm}, then after identifying the total action as the free energy we find
\begin{align}
\begin{split}
S_{EE}=-\frac{\partial F}{\partial T}= - 4 a\log (\frac{f_{0}}{\epsilon} )+\cdots
\end{split}
\end{align}
 where $F=I_{tot}T$ is the free energy functional and $a=\frac{\pi^{2} L^3  }{  \lp^3 }$\,. We identify $S_{EE}$ as the entanglement entropy and is precisely what arises as the log term in the RT prescription. Notice that in this way of calculating the entropy, the power law divergences have cancelled out. We have also checked that the results of Gauss-Bonnet gravity \cite{higher} are reproduced using this approach \cite{prep}.

At this stage one can give an argument as to why the area minimization prescription of RT leads to the same result as above. In an Euclidean path integral, the total action considered above would be a functional of $f(z)$. The way that $f(z)$ gets fixed in the path integral is as follows. In the total action considered above, $f(z)$ appears in $I_{bulk}+I_{GH}^{r=f(z)}+I_{ct}^{r=f(z)}$.  If we consider varying $f(z)$ then since in $I_{bulk}+I_{GH}^{r=f(z)}$, such variations correspond to variations in the $g_{zz}$ component of the metric, these will vanish on using the equations of motion for the background. However, $\delta I_{ct}^{r=f(z)}/\delta f(z)$ has to be set to zero independently. Thus we are led to the minimization prescription. The fact that $T_{tt}=0$ gave rise to the same equation as what follows from the minimization is a consistency check of this argument.
 This argument is similar to the one used by Fursaev \cite{fursaev} except that in our approach we did not have conical singularities in the bulk.  This points to the possibility that there may be methods that do not rely on the replica trick for computing entanglement entropy.

Now one can ask if this observation can be used to extract $f(z)$ from the field theory. Firstly note that in the holographic calculation we put $r=f(z)$ and the resulting slice was a 4d one. We will consider field theory on this 4d slice--the idea is to see if we can construct $f(z)$ by demanding the vanishing of the $tt$ component of the field theory stress tensor in this geometry. This stress tensor is not the usual stress tensor that is evaluted on the $z=0$ surface. We can give a heuristic motivation for our calculation as follows (see \cite{prep} for some more details). It has been argued in \cite{swingle} that there is a connection between the entanglement renormalization scheme of MERA and the way that entanglement entropy is calculated in AdS/CFT. In this connection, the more coarse grained one makes the quantum system, the deeper one is in the IR. The coarse graining is done in a specific way using unitary operators. As such if one starts with the ground state, we expect to be in the ground state after any number of coarse-graining steps with respect to a new hamiltonian that is a unitary transformation of the orginal hamiltonian. Imposing $T_{tt}=0$ along the radial evolution in AdS/CFT will ensure that we are in the ground state. Roughly speaking this is the reason why we are interested in doing the field theory calculation in this seemingly unusual way.

 It is well known that when one considers the expectation value of the stress tensor of any quantum field theory on a curved background \cite{bd}, then there are UV divergences. These UV divergences depend on the local geometry and involve the local Riemann tensor and its contractions. This is expected since the divergences arise due to short wavelength modes which probe local geometry. Crucially the geometric feature of the divergences is independent of the global features of the spacetime as well as the actual quantum state involved \cite{bd, luty}. Once we regularize, the finite stress tensor will depend not only on geometric pieces but also the long wavelength features such as the global properties of the manifold as well as the actual quantum state involved. We will focus on cases where the stress tensor has divergences since here they are governed (upto overall constants) by local geometry \cite{othercase}. This is also to be consistent with the fact that in the holographic calculation we worked with the unrenormalized stress tensor. Let us begin by considering for definiteness a massless scalar whose divergent stress tensor in dimensional regularization about $d=4$ is given by \cite{bd, bunch, error}
\be\label{Ttt}
\langle T_{ab} \rangle_{div}=-\frac{1}{(4\pi)^{d/2}}\frac{1}{d-4}\frac{1}{30}\left(^{(2)}H_{ab}-\frac{1}{3}{} ^{(1)}H_{ab} \right)\,,
\ee
where, 
\begin{align}
\begin{split}
^{(1)}H_{ab} &=-2\nabla_{a}\nabla_{b}\R+h_{ab}(2\Box\R-\frac{1}{2}\R^2)+2 \R \R_{a b }\,   \\ ^{(2)}H_{ab}&=\Box \R_{ab} -2 h^{cd} \nabla_{d}\nabla_{b}\R_{a c}+\frac{1}{2}h_{ab }( \Box\R - \R_{cd } \R^{cd })+2 \R_{a}^{c } \R_{b c }\,. \\  
\end{split}
\end{align}

Following the holographic calculation, if we demand $
\langle T_{tt} \rangle_{div}=0$, then we find \cite{mathe} that for the sphere case, $f(z)=\sqrt{f_0^2-z^2}$ while for the cylinder case the expansion agrees with what arises from the RT prescription upto $O(z^2)$. Specifically,  for the  sphere, when evaluated on-shell $
\R_{ab}=-\frac{3}{L^{2}}h_{ab} \,.
$
Using this it is easy to see that  $^{(1)} H_{ab}$ and $^{(2)}H_{ab}$ will vanish.
For the cylinder we first assume a $f(z)$ of the form, $f(z)= f_{0}+f_{2} z^{2}.$ Then\,,
\be
^{(2)}H_{tt}-\frac{1}{3}{} ^{(1)}H_{tt}=\frac{8 f_2^2 (1+4 f_0 f_2)}{3 f_0} z^2+O(z^4)\,,
\ee
Setting the $O(z^2)$ term to zero gives ($f_2\neq 0$ as otherwise it leads to inconsistency at the next order),
\be
 f_{2}=-\frac{1}{4 f_{0}}
\,,\\
\ee
which is the same as what follows from the RT prescription.
 In the cylinder case, we note that it is precisely upto $O(z^2)$ order that is needed to extract the universal $\log$ term in the entanglement entropy while $O(z^4)$ terms gave rise to subleading non-universal terms. We find that the agreement for the cylinder case $f(z)$ breaks down at $O(z^4)$. This is not surprising since the $O(z^4)$ terms will  also be sensitive to the finite pieces of the stress tensor. What is perhaps surprising is the exact agreement for the sphere.

Let us conclude with some observations. First, we found that the RT area functional arises as a counterterm to get a regularized stress tensor on the $r=f(z)$ slice. As a strong cross-check this result extends easily to Gauss-Bonnet gravity and reproduces the area functional considered in \cite{higher, prep2}. Wald entropy evaluated on the entangling surface gives wrong results for the cylinder case as noted in \cite{higher}. The difference is due to extrinsic curvature terms in the Wald entropy.  Our derivation explains why the area functional does not have terms depending on the 3-extrinsic curvatures as observed in \cite{higher}. The reason is that we want a well defined Dirichlet problem and the presence of 3-extrinsic curvatures in the counterterm action would lead to variations in directions normal to the surface. This way of thinking also suggests a systematic way of deriving the area functional in more complicated examples. For example, while we have shown that the RT functional arises as a counterterm at a CFT fixed point, our approach would also suggest some differences from the RT prescription in the context of RG flows in case there are contributions to the co-dimension two counterterm action from the matter inducing the flow. It would be interesting to study examples where this happens. Second, we have tried to extract information about the way the entangling surface extends into the bulk by using the UV divergences in the field theory. We assumed a background that is given by the $r=f(z)$ slice of AdS. As stated earlier,  ideally, we would like to start by considering $r=f_0$ and a UV cutoff and see how $r$ changes as we change the cutoff--this is an important future problem.   In fact, in the field theory calculation, although we used a massless scalar for concreteness, the same result would hold for any massless field \cite{bd}. This strongly suggests that the form for $f(z)$ at least upto $O(z^2)$ is independent of the actual gravity dual to conformal field theories. This is borne out in the higher derivative calculations with Gauss-Bonnet terms performed in \cite{higher}. Further this procedure was crucially tied to even dimensions since in odd dimensions the analogous divergences are absent in dimensional regularization. In odd dimensions, one would need to use the regularized stress tensor which would then become sensitive to global properties of spacetime as well as the state being used. It will be interesting to extend this to $d=3$ \cite{kleba}.   It will also be interesting to see if there is any connection with the covariant generalizations of the RT proposal \cite{mukund}.

Curiously, eq.(\ref{Ttt}) is what would arise from the `pole' type counterterms in AdS/CFT. These encode the conformal anomaly. In fact the so-called dilaton action \cite{zohar} can be derived by carefully regularizing the pole term \cite{bhss}. This leads us to suspect that it may be possible to derive information about $f(z)$ from the dilaton action itself by appropriately identifying the dilaton with $f(z)$. Another point that we wish to emphasise is that constructing $f(z)$ from entanglement renormalization is a part of the program \cite{swingle} for connecting AdS with continuous Multiscale Entanglement Renormalization Ansatz (cMERA). The fact that we have been able to get some information of $f(z)$ using a field theory calculation gives credence to the RT prescription as well as hope that $f(z)$ construction from AdS/cMERA should be possible.

{\bf Acknowledgments} : We thank Sayantani Bhattacharyya,  Janet Hung, Chethan Krishnan, Gautam Mandal,  Rob Myers, Miguel Paulos, Suvrat Raju and Tadashi Takayanagi  for useful discussions. We especially thank Rob Myers, Miguel Paulos and Tadashi Takayanagi for useful comments on the draft. AS acknowledges support from a Ramanujan fellowship, Govt. of India.


\begin{thebibliography}{99}
\bibitem{odd}
In odd dimensional theories, typically the $\log$ term is missing and is replaced by a constant which is conjectured to be universal.
\bibitem{motiv} Our original motivation for looking at the $tt$ component of the stress tensor came from Jacobson's pioneering work \cite{jacob} connecting thermodynamics to equations of motion in general relativity.
\bibitem{othercase}
 In those cases where the divergences cancel (e.g., for $d=2$ massless scalar or if we consider the entangling surface to be a slab in which case the Weyl tensor vanishes irrespective of what $f(z)$ we choose) it may still be possible to extract $f(z)$ from the finite pieces but now details of the actual quantum state will enter and complicate matters.
\bibitem{error}
Beware! There are sign errors in the expressions for $^{(1)}H_{ab}$ and $^{(2)}H_{ab}$ in \cite{bd}.
\bibitem{mathe} Mathematica notebooks for this can be made available upon request.
\end{thebibliography}
\end{document}